\documentclass[aps,secnumarabic,nobalancelastpage,amsmath,amssymb,
nofootinbib]{revtex4}

\usepackage{graphics}      
\usepackage{graphicx}      
\usepackage{longtable}     
\usepackage{url}           
\usepackage{bm}            
\usepackage{hyperref}
\usepackage[dvipsnames]{xcolor}

\begin{document}

\def\lsim{\:\raisebox{-0.5ex}{$\stackrel{\textstyle<}{\sim}$}\:}

\title{The Effects of QCD Equation of State on the Relic Density of WIMP Dark Matter}
\author         {Manuel Drees}
\email          {drees@th.physik.uni-bonn.de}
\affiliation    {Bethe Center for Theoretical Physics and Physikalisches
  Institut, Universit\"at Bonn,\\Nussallee~12, D-53115 Bonn,
  Germany}
  
\author         {Fazlollah Hajkarim}
\email          {hajkarim@th.physik.uni-bonn.de}
\affiliation    {Bethe Center for Theoretical Physics and Physikalisches
  Institut, Universit\"at Bonn,\\Nussallee~12, D-53115 Bonn,
  Germany}
  
\author         {Ernany Rossi Schmitz}
\email         {ernany@th.physik.uni-bonn.de}
\affiliation    {Bethe Center for Theoretical Physics and Physikalisches
  Institut, Universit\"at Bonn,\\Nussallee~12, D-53115 Bonn,
  Germany}
\date{\today}

\begin{abstract}

  Weakly Interactive Massive Particles (WIMPs) are the most widely
  studied candidate particles forming the cold dark matter (CDM) whose
  existence can be inferred from a wealth of astrophysical and
  cosmological observations. In the framework of the minimal
  cosmological model detailed measurements on the cosmic microwave
  background by the PLANCK collaboration fix the scaled CDM relic
  density to $\Omega_{c}h^2=0.1193\pm0.0014$, with an error of less
  than 1.5\%. In order to fully exploit this observational precision,
  theoretical calculations should have a comparable or smaller
  error. In this paper we use recent lattice QCD calculations to
  improve the description of the thermal plasma. This affects the
  predicted relic density of ``thermal WIMPs'', which once were in
  chemical equilibrium with Standard Model particles. For WIMP masses
  between 3 and 15 GeV, where QCD effects are most important, our
  predictions differ from earlier results by up to $9\% \ (12\%)$ for
  pure $S-$wave ($P-$wave) annihilation. We use these results to
  compute the thermally averaged WIMP annihilation cross section that
  reproduces the correct CDM relic density, for WIMP masses between
  0.1 GeV and 10 TeV.
\end{abstract}

\maketitle

\section{Introduction}
\label{sec:introduction}

Assuming Newtonian gravity, and its extension into General Relativity,
describes gravitational forces correctly, astronomical and
cosmological observations show that most of the matter in our Universe
is a non-luminous, neutral form of matter, called dark matter
(DM). Quantitatively, it accounts for $\sim85\%$ of all matter
\cite{davis1985,PLANCKresults,bennettWMAP}. There are many possible DM
candidates \cite{dg}. Among those, weakly interactive massive
particles (WIMPs) have been studied in most detail. There are several
reasons for the popularity of WIMPs. If they once were in full
chemical equilibrium with the particles of the Standard Model (SM),
which is true if the largest temperature after the most recent period
of entropy production exceeded about 5\% of the WIMP mass, the WIMP
relic density can be calculated unambiguously for a given cosmological
model, independent of initial conditions, using only particle physics
quantities (masses and couplings, or cross sections) as input. This
calculation yields approximately the observed relic density for
roughly weak--strength WIMP annihilation cross sections. This not only
hints at a deep connection between DM and extensions of the SM
addressing some of its shortcomings
\cite{ellis1984,kane1994,edsjo1997,birkedal2006,barger2009}, it also
allows various ways in which the existence of WIMP DM can be probed.
Unfortunately these probes have so far not yielded an unambiguous
signal. Successful WIMP candidates must therefore not only satisfy the
relic abundance constraint, but also constraints coming from direct
\cite{XENON1002012,superCDMS2014,LUX2014} and indirect \cite{Fermi-LAT
  3 analysis,Fermi-LAT 4,Ackermann:2013yva,Fermi-LAT new,WMAP+ACT 3} detection
experiments.

In standard cosmology it is assumed that the comoving entropy density 
remained constant since the epoch when WIMPs were in full thermal
equilibrium with SM particles. The quantity of interest is then the
ratio $Y_\chi$ of the WIMP number density $n_\chi$ and the entropy
density $s$. In order to compute the temperature dependence of
$n_\chi$, one thus needs to know the precise temperature dependence of
$s$. In addition, the expansion rate, described by the Hubble
parameter $H$, is proportional to the square root of the total energy
density $\rho$; hence the temperature dependence of $\rho$ also has to
be known. In early analyses \cite{ellis1984,kt} $s(T)$ and $\rho(T)$
were calculated ignoring all interactions between SM particles,
i.e. treating the thermal plasma as a relativistic free gas. This is
quite a good approximation for most temperatures. However, it was
realized early on \cite{kt,Gondolo_Gelmini} that this approach fails
for temperatures near the deconfinement transition, i.e. the
transition from hadronic (pions, kaons, \dots) to partonic (quarks,
gluons) degrees of freedom. More recently it was realized that also
for some range of temperatures above this transition the interactions
between quarks and gluons should not be ignored
\cite{Hindmarsh,Laine_Schroeder}. 

In this paper we carefully model the effect of strong interactions on
the temperature dependence of the energy and entropy densities. We
exploit recent calculations performed in the framework of the Hadron
Resonance Gas (HRG) model well below the deconfinement transition
\cite{HRG model}, smoothly matching to results from lattice QCD (LQCD)
at $T=100$ MeV \cite{Bazavov:2014pvz}. We find that this can change
the predicted relic density by more than 5\% compared to earlier
treatments \cite{Gondolo_Gelmini,Hindmarsh,Laine_Schroeder} of strong
interaction effects on the thermal plasma.

We then use our improved treatment of strong interaction effects to
determine the value of $\langle \sigma v \rangle$, the thermally
averaged product of WIMP annihilation cross section times relative
velocity of the annihilating WIMPs, that is required to reproduce the
correct relic density, under the assumption that $\langle \sigma v
\rangle$ is independent of temperature. As already noted in
ref.~\cite{Steigman}, depending on the WIMP mass, this product can
deviate by up to a factor of $\sim 1.5$ in either direction from the
``canonical'' value of $3 \cdot 10^{-26}$ cm$^3$s$^{-1}$. This is
important since experiments are now beginning to be sensitive to the
canonical cross section. For example, analyses of FERMI-LAT gamma ray
observations of nearby dwarf galaxies and of the diffuse gamma ray
emission in the galaxy have, for some combinations of WIMP masses and
WIMP annihilation final states, constrained the WIMP annihilation
cross section to be lower than the canonical one \cite{Fermi-LAT 3
  analysis,Fermi-LAT 4,Ackermann:2013yva}.

The remainder of this paper is organized as follows. In
Sec.~\ref{sec:secondsection} we review the calculation of the relic
abundance of thermal WIMPs. We outline earlier estimates of $s(T)$ and
$\rho(T)$, and explain in detail how we compute these quantities.  In
Sec.~\ref{results} we present our result for $\left\langle \sigma v
\right\rangle$, and compare with previous results. In
Sec.~\ref{constraints} we compare our $\left\langle \sigma v
\right\rangle$ with bounds on this quantity from indirect WIMP
detection experiments and CMB anisotropies. Finally, we summarize the
main results of our work in Sec.~\ref{conclusion}.

\section{Calculation of the Relic Abundance}
\label{sec:secondsection}

\subsection{Basic Framework}
\label{sub:secondfirstsubsection}

The starting point of our calculation is the Boltzmann equation, which
describes the time evolution of the number density $n_\chi$ of a
stable particle species $\chi$ that can annihilate into a pair of SM
particles \cite{kt}:
\begin{equation} \label{boltz}
\frac {dn_\chi} {dt} + 3 H n_{\chi} = -\langle \sigma v \rangle \left(
  n_{\chi}^{2} - n_{\chi,\textrm{eq}}^{2} \right) \,.
\end{equation}
Here $t$ is the cosmological time, $H$ is the Hubble parameter,
$\langle \sigma v \rangle$ is the thermal average of the product of
WIMP annihilation cross section and M$\o$ller velocity\footnote{In the 
non--relativistic limit the M$\o$ller velocity becomes the relative 
velocity between the annihilating WIMPs.}, and
$n_{\chi,\textrm{eq}}$ is the density of $\chi$ particles
in thermal equilibrium. We have assumed that $\chi$ particles are
self--conjugate (Majorana) particles. With minor modifications our
results can also be applied to particles that are not self--conjugate,
such as Dirac fermions or complex scalar particle. In this case one
will in general have to track the densities of particles and
antiparticles separately.

Since the total entropy is assumed to be conserved, $S = sa^3
=$const. where $a$ is the scale factor in the
Friedman--Robertson--Walker metric and $s$ is the entropy density, it
is convenient to define $Y_\chi \equiv n_{\chi}/s$. Using the
definition $H = (1/a) da / dt$ it is easy to see that this definition
absorbs the dilution term $3 H n_\chi$ on the right--hand side of the
Boltzmann eq.(\ref{boltz}).

Moreover, since $s$ as well as $n_{\chi,{\rm eq}}$ depend explicitly
on temperature rather than on time, one replaces time $t$ by $x\equiv
m_{\chi}/T$. To that end, the entropy density is written as
\begin{equation} \label{def_h}
s(T) = \frac{2 \pi^2} {45} h(T) T^3\,.
\end{equation}
Here $h(T)$ counts the effective number of relativistic degrees of
freedom (d.o.f.) that contribute to $s$, i.e. if all relevant
particles have a common temperature $T$ and masses $\ll T$, $h(T)
\rightarrow \sum_i {\bf g}_i$, where ${\bf g}_i$ is the number of
effective internal degrees of freedom for particle species $i$. For
example, a massless photon contributes ${\bf g}_\gamma = 2$, a
massless Dirac fermion (with both helicities being in thermal
equilibrium) contributes ${\bf g}_e = 4 \cdot 7/8 = 3.5$, and so
on. Using $\frac {d} {dt} \left[ h(T) T^3 a^3 \right] = 0$ one then
finds
\begin{equation} \label{tofT}
\frac{dT} {dt} = - \frac {TH} { 1 + \frac {1}{3} \frac {d \, \ln h(T)}
  {d \, \ln T} }\,.
\end{equation}

We thus also need to know the temperature dependence of the Hubble
parameter. To that end we use the Friedmann equation $H^2 = 8\pi G_N
\rho/3$, where $\rho$ and $G_N$ are the energy density and Newton's
gravitational constant, respectively. During the radiation dominated
epoch, in which the decoupling of WIMPs falls, the energy density can
be written as
\begin{equation} \label{def_g}
\rho(T) = \frac{\pi^2} {30} g(T) T^4\,.
\end{equation}
Here $g(T)$ counts the effective number of relativistic d.o.f.  that
contribute to $\rho$. In the limit where all particle masses can be
ignored, $g(T) = h(T)$, i.e. the numbers of d.o.f. defined via the
entropy density and via the energy density are the same, but in
general this is not the case.

Putting everything together, we have \cite{Srednicki,Gondolo_Gelmini}
\begin{equation} \label{boltz1}
\frac {dY_\chi} {dx} = \lambda g_*^{1/2} \frac{1}{x^2} \left( Y_{\chi,
    {\rm eq}}^{2} - Y_\chi^{2} \right)\,, 
\end{equation}
where we have introduced
\begin{equation}  \label{gstar}
g_{*}^{1/2} = \frac {h} {g^{1/2}} \bigg[1 + {\frac{1}{3}}{d({\rm ln}h)
  \over \,d({\rm ln}T)}\bigg]
\end{equation}
and
\begin{equation} \label{lambda}
\lambda = \sqrt{ \frac {\pi} {45 G_N} } m_\chi \langle \sigma v
\rangle\,,
\end{equation}
$m_\chi$ being the WIMP mass. Numerically, $\lambda \equiv 2.76 \times
10^9$ for a WIMP mass of 1 GeV and $\langle \sigma v \rangle =
10^{-26}$ cm$^3$s$^{-1}$. The quantities $g$ and $h$ appearing in
eq.(\ref{gstar}) are the effective number of relativistic
d.o.f. defined via the energy density, eq.(\ref{def_g}), and entropy
density, eq.(\ref{def_h}), respectively. Note that we need to know
both $g$ and $h$: both appear in the definition (\ref{gstar}) of
$g_*^{1/2}$, and $h$ also appears in the explicit expression for the scaled
equilibrium density $Y_{\chi,{\rm eq}}$. WIMPs decouple when they are
essentially non--relativistic, so we can write
\begin{equation} \label{yeq}
Y_{\chi,{\rm eq}}(x) = {n_{\chi,{\rm eq}} \over s} = {45\over 2\pi^4}
\bigg( {\pi\over 8} \bigg)^{1/2} {{\bf g}_\chi \over h}\,
x^{3/2}{\rm exp}(-x),
\end{equation}
where ${\bf g}_{\chi} = 2$ for a neutral Majorana fermion.

\subsection{The Functions $g(T)$ and $h(T)$}

Most papers on the calculation of the WIMP relic density focus on the
annihilation cross section. Our concern is instead an accurate
calculation of the effective numbers of degrees of freedom encoded in
the functions $g(T)$ and $h(T)$. To that end, we have to compute the
energy density $\rho(T)$ and the pressure $p(T)$; the entropy density
is then given by $s(T) = [\rho(T) + p(T)]/T$. If the single particle
distribution functions $f_i(\vec{k}, T)$ for a particle species $i$ is
known, the contribution of this species to the energy density and
pressure are given by \cite{kt} $\rho_i(T) = {\bf g}_i/(2\pi)^3 \cdot
\int d^3 k E_i(\vec k) f_i(\vec k, T), \ p_i(T) = {\bf g}_i/(2\pi)^3
\cdot \int d^3k (\vec k)^2 f(\vec k, T) / (3 E_i(\vec k))$; here ${\bf
  g}_i$ is again the number of degrees of freedom of species $i$, and
$\vec k$ is the three--momentum.

The simplest case is that of a free (non--interacting) particle with
mass $m_i$. The distribution function is then the Bose--Einstein or
Fermi--Dirac distribution function, which depends only on the ratio of
the energy $E_i = \sqrt{m_i^2 + \vec k^2}$ and the temperature.
Introducing the dimensionless quantities $y_i = E_i / m_i$ and $x_i =
m_i/T$, one has \cite{kt,Gondolo_Gelmini}: 
\begin{eqnarray}
g_i(T) &=& \frac{15\bf{g}_i} {\pi^4} x_i^4 \int_{1}^{\infty} \frac
{y_i^2 \sqrt{y_i^2 - 1} } { \exp(x_iy_i) \pm 1 } dy_i \,; \label{gfree} \\
h_i(T) &=& \frac{45\bf{g}_i} {4\pi^4} x_i^4 \int_{1}^{\infty}
\frac{ \sqrt{ y_i^2 -1 } } { \exp(x_iy_i) \pm 1 } \frac{4y_i^2-1}{3} dy_i \,.
\label{hfree}
\end{eqnarray}
In the denominators of the integrals $+1$ applies to fermions and $-1$
to bosons.

In the presence of strong interactions, Eqs.(\ref{gfree}) and
(\ref{hfree}) no longer provide good approximations. Before describing
our own calculation of these functions, we briefly review the state of
the art, as encoded in the widely used program packages for the
calculation of the WIMP relic density {\tt DarkSUSY}
\cite{Gondolo:2004sc}, {\tt micrOMEGAs} \cite{Belanger:2013oya} and
{\tt SuperIso} \cite{Arbey:2011zz}. 

It was realized quite early on that care has to be taken to describe
the thermodynamics of the early universe around the deconfinement
transition. In \cite{Olive alone} the interactions between hadrons and
between partons were approximated by simple non--relativistic
potentials. Ref.~\cite{Olive et al.} instead used free particles, and
defined the transition temperature from the hadronic to the partonic
phase as that temperature where the two calculations give the same
entropy density; note that the hadronic phase gives a quickly rising
$h(T)$, since the number of contributing hadrons quickly increases
with temperature. This ``hadron resonance gas model'' was used in all
subsequent calculations at sufficiently low temperatures, including
our own.

One problem of the simple definition of the transition temperature
used in ref.\cite{Olive et al.} is that it leads to a discontinuity in
$g(T)$. Ref.\cite{Srednicki} therefore used smooth functions
interpolating between the hadronic and partonic phases. While these
functions ensure that not only $g(T)$ and $h(T)$, but also their
derivatives are smooth, they were not based on dynamical
considerations. The authors advocated estimating the uncertainty by
using two quite different values, 150 and 400 MeV, for the transition
temperature. The same functions were used in
Ref.\cite{Gondolo_Gelmini}; the functions for a transition temperature
of 150 MeV are still used by default in the computer packages
mentioned above.

The first attempt to include the results of lattice QCD calculations
was due to Hindmarsh and Philipsen \cite{Hindmarsh}. At that time the
most accurate lattice QCD calculations did not include dynamical
quarks. There was some evidence that the ratio of the true pressure to
the corresponding value for non--interacting particles shows little
dependence on the number of quark flavors \cite{Karsch_et_al}.
Ref.\cite{Hindmarsh} therefore scaled the contribution of all strongly
interacting partons by the same correction function, determined from
pure glue lattice calculations \cite{Karsch_et_al}; at $T = 1.2$ GeV,
these were matched to perturbative calculations
\cite{Kajantie_et_al}.

The treatment by Laine and Schroeder \cite{Laine_Schroeder} is rather
similar. However, their results are based on a different set of pure
glue lattice QCD calculations \cite{Laine_lattice}. Moreover, they
match to perturbative calculations at the much lower temperature of
350 MeV. Finally, they include the quark mass dependence up to
next--to--leading order, ${\cal O}(g^2)$, in the perturbative
expansion. In particular, they point out that charm quarks make
non--negligible contributions already at temperatures of a few hundred
MeV.

We now describe own treatment. At temperatures well above the
electron mass, we treat all SM particles without strong interactions
as free particles, i.e.  we use Eqs.(\ref{gfree}) and (\ref{hfree})
for these particles. This includes the leptons, the electroweak gauge
bosons as well as the single physical Higgs boson of the SM. Note
that for the physical Higgs mass, $m_H \simeq 125$ GeV, in the
Standard Model electroweak symmetry breaking leads to a smooth
cross--over, not a phase transition
\cite{Kajantie:1996mn,Csikor:1998eu,Fodor:1999at}; hence the comoving
entropy density remains constant, as assumed in the derivation of
eq.(\ref{boltz1}). For simplicity we compute $g(T)$ and $h(T)$ using
free, massive $W^\pm$ and $Z$ bosons (with three d.o.f. each) and a
single physical Higgs boson even for temperatures above the
electroweak cross--over, where a more accurate treatment would use
massless gauge bosons (with two d.o.f. each) and a massive complex
Higgs doublet (with four d.o.f.). The difference between these
treatments is very small, and will only affect the relic density of
WIMPs with masses above 2 TeV.

The main focus of our work is on the effect of QCD interactions. These
are most important around the deconfinement transition, which was also
a smooth crossover \cite{Aoki:2006we} with conserved total entropy. We
use the results of a recent lattice calculation with $N_f = 2+1$
active flavors (meaning equal masses are used for $u$ and $d$ quarks,
but the larger mass of the strange quark has been taken into account)
\cite{Bazavov:2014pvz}. This calculation covers temperatures between
100 and 400 MeV. Ref.\cite{Bazavov:2014pvz} provides a parameterization
of the pressure due to $u,d,s$ quarks and gluons in this temperature
range:
\begin{equation} \label{eosfit}
\frac{p}{T^4} =
\frac{1}{2} \left[ 1 + \tanh \left(c_t (\bar{t} -t_0) \right) \right] \cdot
\frac{p_{\rm id} +  a_n/\bar{t} + b_n/\bar{t}^2 + d_n/\bar{t}^4 } { 1
  + a_d/\bar{t} + b_d/\bar{t}^2 + d_d/\bar{t}^4} \,.
\end{equation}
Here $\bar{t} = T/T_c$, $T_c=154$~MeV being the QCD transition temperature.
In this parameterization, $p_{\rm id}= 19 \pi^2/36$ is the ideal gas value of
$p/T^4$ for QCD with three massless quarks. The values of the
numerical coefficients appearing in eq.(\ref{eosfit}) are listed in
Table~\ref{tab:eosfit}. Recall that $\tanh(z)$ approaches unity for
large argument $z$. Therefore eq.(\ref{eosfit}) automatically
approaches the ideal gas value for $T \gg T_c$, i.e. $\bar t \gg
1$. Moreover, ref.\cite{Bazavov:2014pvz} shows that eq.(\ref{eosfit})
matches quite well to available perturbative calculations at higher
temperature. We therefore use this parameterization to describe the
contribution from $u,d,s$ quarks and gluons for all temperatures above
100 MeV. 
\begin{table}[h] \label{tab:eosfit}
\begin{center}
\begin{tabular}{|c|c|c|c|}
\hline
$c_t$&$a_n$&$b_n$&$d_n$ \\
3.8706&-8.7704&3.9200&0.3419 \\
\hline
\hline
$t_0$&$a_d$&$b_d$&$d_d$ \\
0.9761&-1.2600&0.8425&-0.0475 \\
\hline
\end{tabular}
\end{center}
\caption{Parameters used in eq.~(\ref{eosfit}) to describe the pressure of
  (2+1)--flavor QCD.} 
\end{table}

Once the pressure $p$ is known, the energy density $\rho$ can be
computed from the relation between the trace of the energy--momentum
tensor, also called the trace anomaly, and the pressure
\cite{Bazavov:2014pvz}:
\begin{equation} \label{traceanomaly}
\frac{I(T)}{T^4} =\frac{\rho -3p}{T^4}=T\frac{d}{d T}
\left(\frac{p}{T^4}\right) \,.
\end{equation}
Since we have the analytical expression (\ref{eosfit}) for $p/T^4$, we
can easily obtain its derivative, and thus $I(T)$; the first equation
in (\ref{traceanomaly}) then allows to compute $\rho(T)$ from $I(T)$
and $p(T)$. Finally, once $\rho(T)$ and $p(T)$ are known, $s(T) =
[\rho(T) + p(T)]/T$ can also be computed.

As noted above, the effect of the charm quark is not negligible at
temperatures near $T_c$ \cite{Laine_Schroeder}. We included its
contribution to the functions $g$ and $h$ using lattice QCD results
from Table 6 of \cite{Borsanyi:2010cj}, using the physical ratio of
charm and strange quark masses, $m_c/m_s=11.85$ \cite{Davies:2009ih}.
This gives us $p_c/T^4$, from which the charm contributions to $\rho$
and $s$ can be obtained as outlined in the previous paragraph. This
description is valid for $T \leq 1$ GeV. For larger temperatures we
smoothly match to the ideal gas results (\ref{gfree}) and
(\ref{hfree}), using a fit function like (\ref{eosfit}) with $p_{id}=7
\pi^2/60$ and different values for coefficients and powers to
interpolate between the two regimes. This ensures that not only the
functions $g$ and $h$ but also their first derivatives are smooth
everywhere.

Bottom and top quarks contribute significantly only at high
temperatures, where even QCD interactions have become relatively
small. We therefore treat these quarks as free particles, with
on--shell masses given by the Particle Data Collaboration
\cite{PDG_data}.

At temperatures lower than $T_{c}$, the thermodynamic behavior of QCD
can be described by the hadron resonance gas model, in which all the
hadrons and hadron resonances are considered to contribute to the
thermodynamics quantities as non--interacting particles. As noted
above, this has been used already in the early treatments \cite{Olive
  alone, Olive et al.}. Ref.~\cite{Bazavov:2014pvz} shows that for
temperatures between 100 MeV and $T_c$ it matches well to the QCD
results parameterized in eq.(\ref{eosfit}). A convenient
parameterization of the trace anomaly in this model can be found in
ref.~\cite{HRG model}:
\begin{equation}  \label{i_low}
\frac {I(T)} {T^4} = \frac{\rho - 3 p}{T^4} = a_1 T +  a_2 T^3 + a_3 T^4
+ a_4 T^{10} \,, 
\end{equation}
with $a_1 = 4.654$ GeV$^{-1}$, $a_2 = -879$ GeV$^{-3}$, $a_3 = 8081$
GeV$^{-4}$, $a_4 = -7039000$ GeV$^{-10}$. This parameterization is
valid for $70$ MeV $\leq T \leq T_c$. We use it to describe the
contribution from strongly interacting particles for all temperatures
$T < 100$ MeV, using cubic splines to interpolate smoothly to QCD
results at $T > 100$ MeV. The contribution from charmed particles is
negligible at these low temperatures.\footnote{Strictly speaking the
  hadronic contribution to $g$ and $h$ should become exponentially
  small at $T \ll m_\pi = 140$ MeV. At very low temperatures
  eq.(\ref{i_low}) therefore is not accurate. However, this is not
  important for us, since for $T \ll 100$ MeV the hadronic
  contribution is in any case very small; this small contribution need
  not be described very accurately.}

By inverting eq.~(\ref{traceanomaly}), the pressure can be calculated:
\begin{equation} \label{p_int}
\frac{p(T)}{T^4} = \frac{p_0}{T_0^4} + \int_{T_0}^T dT' \frac {I(T')}
{T^{\prime5}} \, , 
\end{equation}
using the numerical result $p(T_0)/T^4_0=0.1661$ at $T_0 = 70$ MeV
\cite{HRG model}. The integral in eq.(\ref{p_int}) can easily be
evaluated analytically if $I(T)$ is given by
eq.(\ref{i_low}). Eq.(\ref{p_int}) thus again provides us with an
analytical parameterization of the pressure, from which we can compute
the energy and entropy densities as described above.

At temperatures below $1$~MeV, the effect of neutrino decoupling
should be included. The rate of reactions changing the $\nu_\mu$ and
$\nu_\tau$ number densities actually becomes smaller than the Hubble
parameter, indicating decoupling, at a temperature of several
MeV. However, at first the expansion of the universe affects photons
and neutrinos in the same way even after neutrino decoupling, i.e. the
photon and neutrino temperatures remain the same. This changes only
once $e^+e^-$ pairs begin to annihilate, at $T \simeq m_e$. Since
neutrinos are already (almost) decoupled by this time, the entropy
that was stored in electrons and positrons gets transferred (almost)
entirely to photons, not to neutrinos. In the limit where neutrino
decoupling was complete when electron decoupling began, this argument
shows that for $T \ll m_e$ the ratio of relic photon and neutrino
temperatures is $T_\gamma/T_\nu=(11/4)^{1/3}$. Actually (electron)
neutrinos were not completely decoupled at $T \simeq m_e$. This can be
described by writing
\begin{equation} \label{hlow}
h = 2\left[ 1 + \frac{7}{8}  \, N_{\rm eff} \left( \frac{4}{11}
\right)^{4/3} \right] \,,
\end{equation}
\begin{equation} \label{glow}
g= 
2\left[ 1 + \frac{7}{8} \, N_{\rm eff} \left( 
\frac{4}{11}\right)\right] \,,
\end{equation}
with $N_{\rm eff} \simeq 3.046$. Note that these expressions include
the contribution from the photon, with ${\bf g}_\gamma =
2$. Eqs.(\ref{hlow}) and (\ref{glow}) are applicable for $T \ll m_e$,
in practice for $T \leq 50$ keV. As mentioned above, for $T > 1$ MeV
we have $T_\nu = T_\gamma$. For $50$ keV $< T < 1$ MeV we use
numerical results from Fig.~1 of \cite{Lesgourgues:2012uu} to
determine the evolution of $T_\nu$ with respect to $T_{\gamma}$.  This
can then be plugged into eqs.(\ref{hlow}) and (\ref{glow}) instead of
$T_\nu/T_\gamma=(4/11)^{1/3}$ to compute the photon and neutrino
contribution to $g(T)$ and $h(T)$. Note that the temperature $T$ is
defined to be that of the photons, $T = T_\gamma$.

\begin{figure}[h] 
\begin{center}
\includegraphics[width=13.0cm]{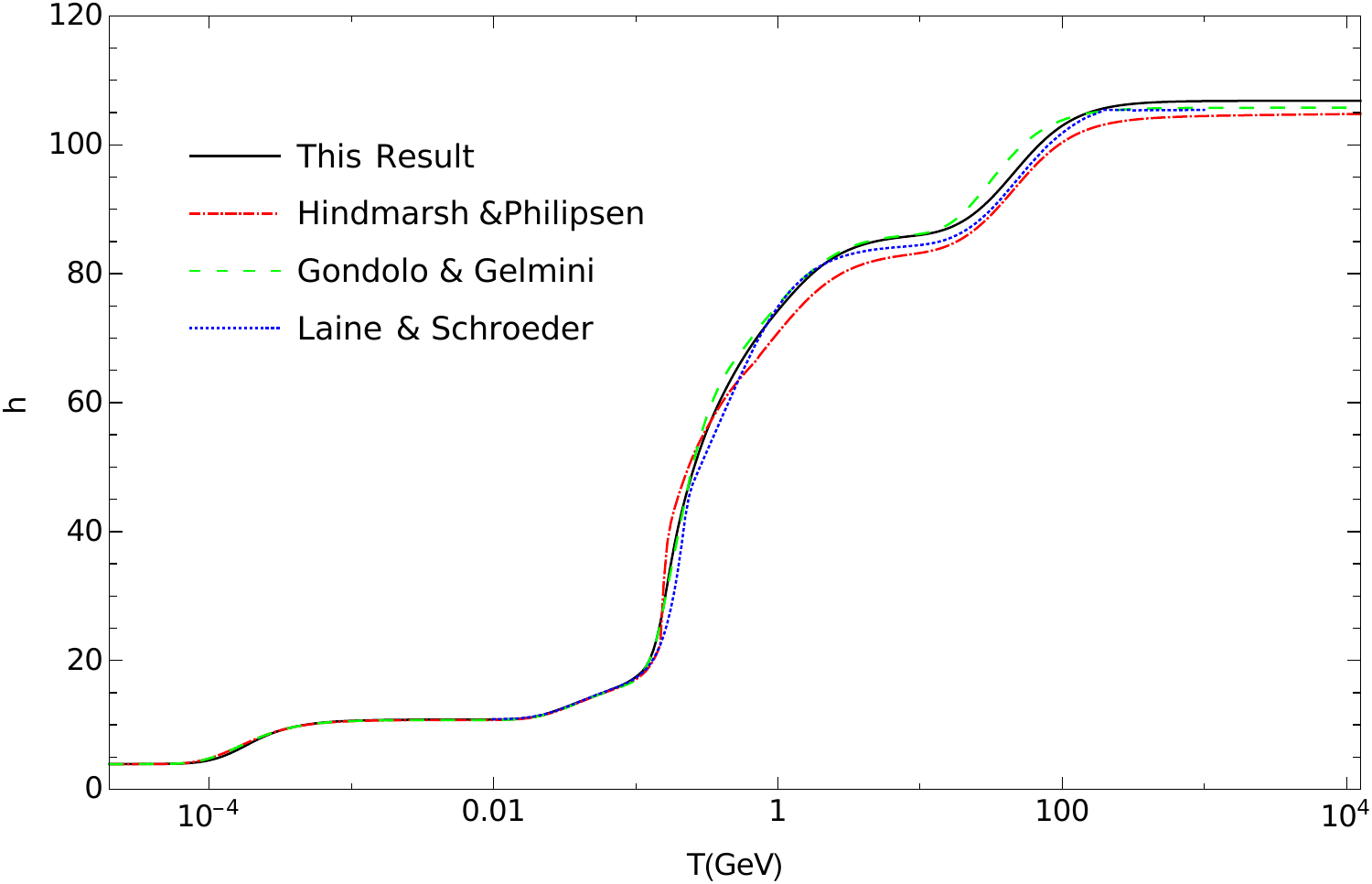}
\includegraphics[width=13.0cm]{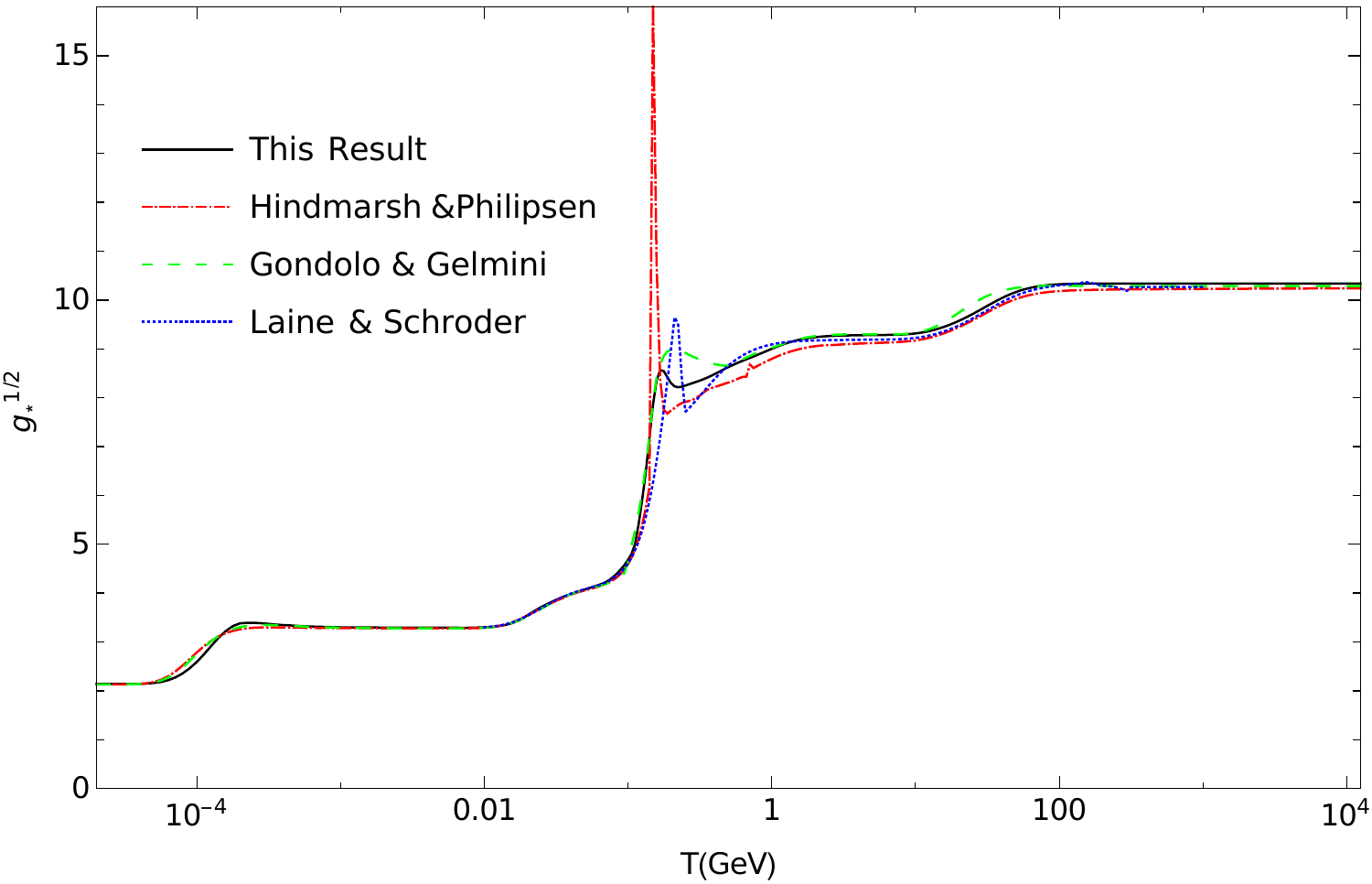}
\end{center}
\caption[] {The functions $h(T)$ (top frame) and $g_*^{1/2}(T)$
  (bottom) defined in eqs.~(\ref{def_h}) and (\ref{gstar}). The
  original calculation by Gondolo and Gelmini \cite{Gondolo_Gelmini},
  based on results from ref.~\cite{Srednicki}, are shown by the green
  dashed curves. The red dot--dashed and blue dotted curves show
  results from refs.~\cite{Hindmarsh} and \cite{Laine_Schroeder},
  which are based on pure glue lattice QCD calculations. The black
  solid curves depict our results, which are based on lattice
  calculations with $N_f = 2+1$ dynamical quark flavors.}
  \label{doffig}
\end{figure}

\section{Results and Comparison with Previous Studies}
\label{results}

We are now ready to present some numerical results.
Figure~\ref{doffig} shows the functions $h(T)$ and $g_*^{1/2}(T)$ that
parameterize thermodynamic effects in the Boltzmann equation
(\ref{boltz1}). The black solid curves show our results, while the
green dashed, red dot--dashed and blue dotted curves show results from
refs.~\cite{Gondolo_Gelmini}, \cite{Hindmarsh} and
\cite{Laine_Schroeder}, respectively.

Since we include the effect of $e^+e^-$ decoupling on the neutrino
background, described by $N_{\rm eff} \simeq 3.046$ in
eq.(\ref{hlow}), we obtained $h\left(T_{\gamma,0}\right) = 3.9387$ in
our calculation; here the present photon temperature $T_{\gamma,0} =
2.7255\pm0.0006$ K \cite{astro}.  Our current value of $h$ is thus
slightly higher than $h\left(T_{\gamma,0}\right)=3.9138$ in
ref.\cite{Hindmarsh} and $h\left(T_{\gamma,0}\right)=3.9139$ in
ref.\cite{Gondolo:2004sc}. Note that for a given value of
$Y_\chi(T_{\gamma,0})$ the final relic density $\Omega_\chi h^2$ is
directly proportional to $h(T_{\gamma,0}) T^3_{\gamma,0}$.

Because both deconfinement of quarks and gluons and the restoration of
the electroweak gauge symmetry are associated with smooth cross--overs
rather than true phase transitions, the functions $g(T)$ and $h(T)$
are smooth everywhere. As noted in the previous Section, we ensured
smoothness of these functions by using cubic splines to interpolate
between different temperature regions. 

There clearly are some differences between the four calculations. These
are most visible near the QCD deconfinement transition. Moreover, the
differences are more visible in $g_*^{1/2}$, largely due to the
derivative term in eq.(\ref{gstar}), which accentuates the differences
between the various treatments. We see that the older calculation
\cite{Srednicki} used in \cite{Gondolo_Gelmini} overestimates
$g_*^{1/2}$ somewhat for $T \simeq 0.1$ GeV, compared to all three
calculations using results from lattice QCD. The treatment of
ref.\cite{Hindmarsh} gives a discontinuity at $T = T_c$, and hence
a divergent derivative, yielding a formally infinite spike in
$g_*^{1/2}$. The continuity has been smoothed out in {\tt micrOMEGAs}
\cite{Belanger:2013oya}, from which we took the numerical results for
$g$ and $h$. We see that this treatment still gives a prominent spike
in $g_*^{1/2}$. Apart from this spike, ref.\cite{Hindmarsh} predicts a
smaller value of $g_*^{1/2}$ in this temperature range than we
do. Finally, the prediction for $g_*^{1/2}$ from
ref.\cite{Laine_Schroeder} is quite close to our own result, except
for some oscillatory behavior just above the QCD transition
temperature.

In order to explore the effect of changes in $h$ and $g_*^{1/2}$ 
on the WIMP relic density, we solve the Boltzmann equation
(\ref{boltz1}) numerically. If we choose the initial value $x_i$ of
$x$ to be $\lsim 10$, the final result is independent of the input
value $Y_\chi(x_i)$. We numerically track the behavior of $Y_\chi(x)$
up to $x = 1,000$, at which point $Y_\chi$ has become practically
constant. The present scaled relic density times squared scaled Hubble
constant can then be computed from
\begin{equation} 
\label{relic}
\Omega_\chi h^2 = {\rho_{\chi} \over \rho_{crit}} h^2 = \frac{ m_{\chi}
  Y_{\chi,0} s_0} {3H_{0}^{2}/8\pi G_N} \left( \frac {H_0}
  {100 \ {\rm km \, s^{-1} \, Mpc^{-1}}} \right)^2
\end{equation}
where $Y_{\chi,0} = Y_\chi(x=1000)$ and $s_0=2891.2$ cm$^{-3}$; the
subscript $0$ denotes quantities evaluated at the present time
\cite{astro}. Note that $H_0$ cancels on the right--hand side of
eq.(\ref{relic}); this is why the relic density is usually quoted in
the combination $\Omega_\chi h^2$. Numerically, $\Omega_\chi h^2 =
2.7889 \cdot 10^8 Y_{\chi,0} m_\chi/(1 \ {\rm GeV})$.

\begin{figure}[h]   
\begin{center}
\includegraphics[width=13.0cm]{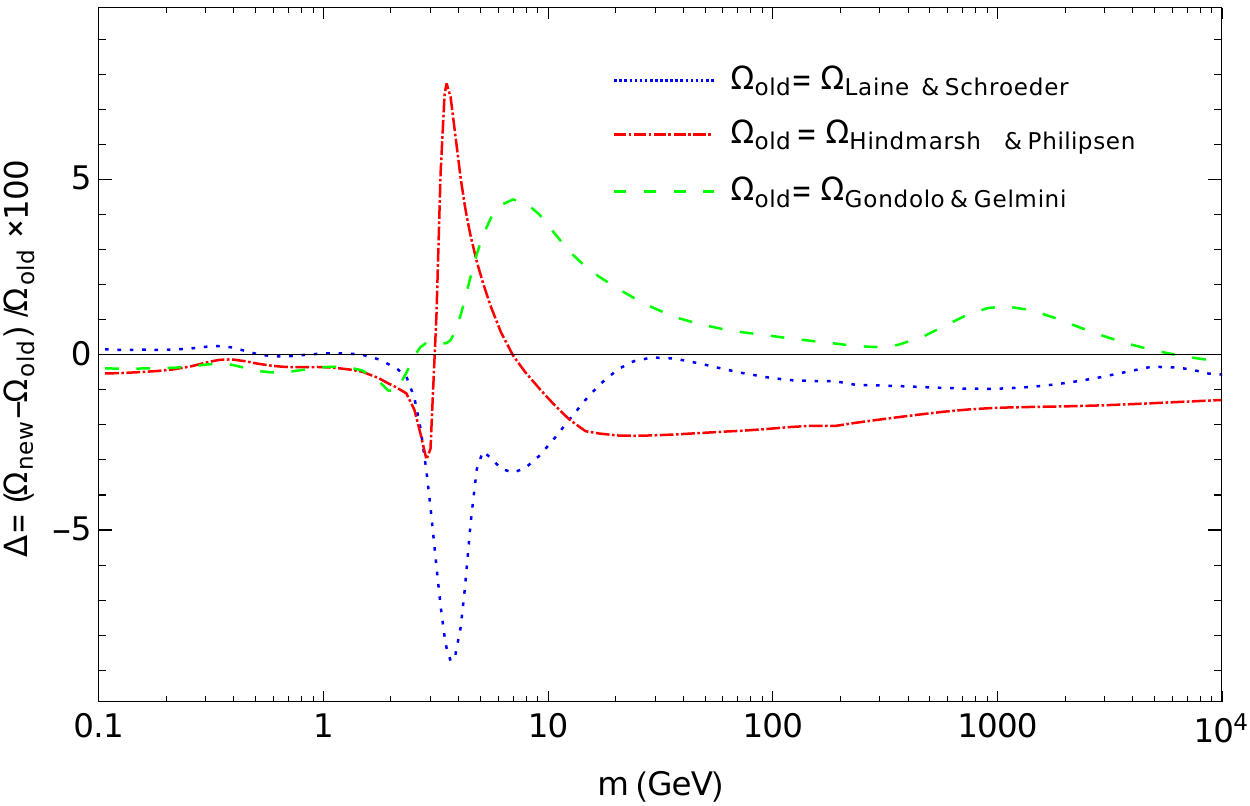}
\includegraphics[width=13.0cm]{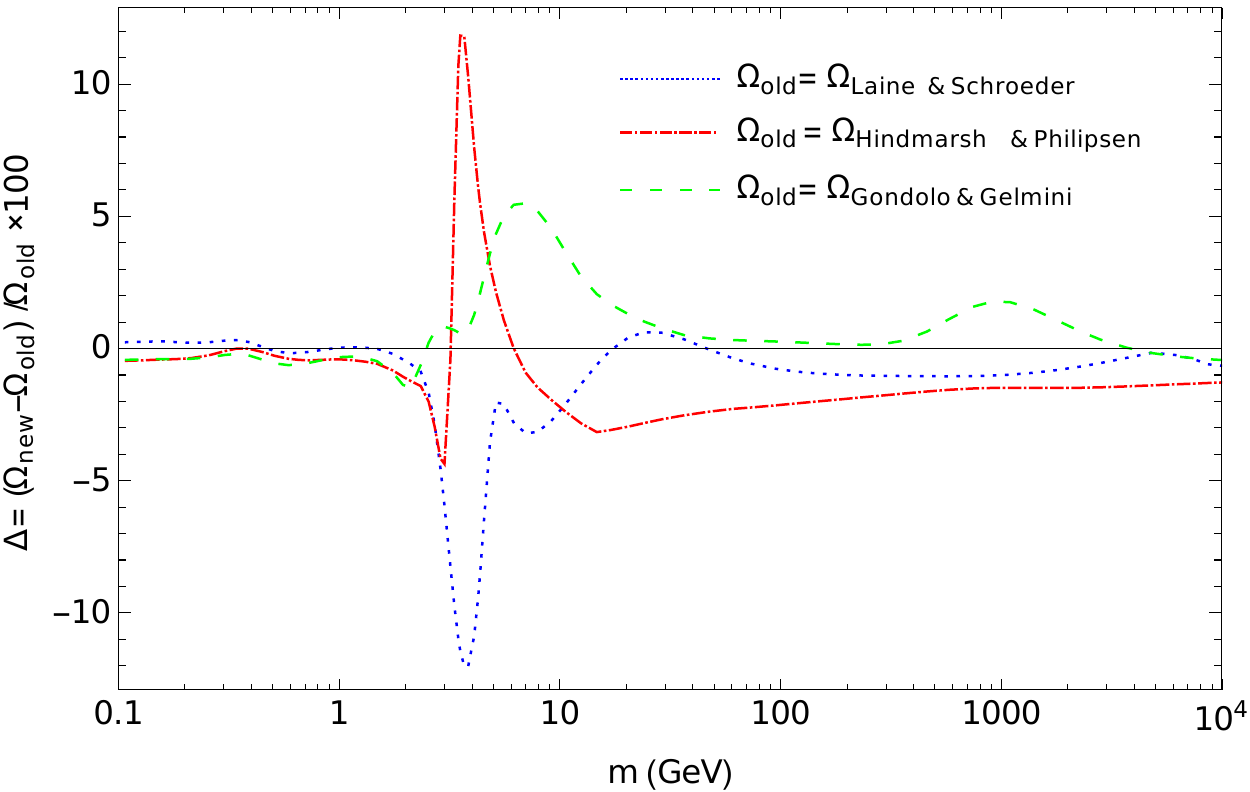}
\end{center}
\caption[] {The relative difference between the predicted relic
  density of a Majorana WIMP between our calculation and a calculation
  using the same older results for the functions $h$ and $g_*^{1/2}$
  shown in Fig.~\ref{doffig}, as function of the WIMP mass. The upper
  frame is for a constant $\langle \sigma v \rangle$, chosen such that
  our prediction for $\Omega_\chi h^2 = 0.1193$, while the lower frame
  is for a pure $P-$wave annihilation, with $\langle \sigma v \rangle
  = 1.2 \cdot 10^{-24}$ cm$^3$s$^{-1} \cdot T/m_\chi$. These results are
  almost independent of the numerical size of the annihilation cross
  section.}
  \label{relicpercentage}
\end{figure}

The change of the predicted WIMP relic density due to our more refined
treatment of the functions $h$ and $g_*^{1/2}$ is illustrated in
fig.~\ref{relicpercentage}. The upper frame shows results for a
temperature independent $\langle \sigma v \rangle$, while in the lower frame
we have assumed $\langle \sigma v \rangle \propto 1/x$. These
behaviors describe the thermally averaged cross section at small
velocity away from poles (i.e. if the WIMPs cannot annihilate into any
particle $\phi$ with $m_\phi \simeq 2 m_\chi$) and thresholds (i.e. if
the WIMPs are significantly heavier than all relevant final--state
particles), if the annihilation occurs from a pure $S-$wave and pure
$P-$wave initial state, respectively. The latter occurs, for example,
for a Majorana WIMP annihilating into light SM fermions, or for a
complex scalar annihilating through $s-$channel exchange of a gauge
boson. Not unexpectedly, we observe the largest differences for WIMP
masses of a few GeV, which decouple just above the QCD transition
temperature. The differences amount to up to $9\%$ for the pure
$S-$wave, and up to $12\%$ for the pure $P-$wave.

The results of fig.~\ref{relicpercentage} can be understood in more
detail using the approximate analytical solution of the Boltzmann
equation developed in ref.\cite{kt}:
\begin{equation} \label{appsol}
Y_{\chi,0} \propto \frac {x_F} {g_*^{1/2}(T_F) \langle \sigma v
  \rangle(T_F) }\,, \ \ \ x_F \propto \ln\left( m_\chi g_*^{1/2}(T_F)
  \langle \sigma v \rangle(T_F) / h(T_F) \right)\,.
\end{equation}
Very roughly, $x_F \sim 20$ for WIMP masses and annihilation cross
sections of interest. This equation shows that $g_*^{1/2}$ affects the
final result more strongly than $h$ does, which appears only
logarithmically.  The derivation of eq.(\ref{appsol}) assumes that
$g_*^{1/2}$ and $h$ are constant around the WIMP decoupling
temperature $T_F = m_\chi / x_F$. This is not a very good
approximation near the QCD deconfinement transition, where these
functions change rapidly, as we saw in fig.~\ref{doffig}. However, we
can see directly from the Boltzmann equation that the most relevant
temperature range is that around the decoupling temperature. At higher
temperatures, $Y_\chi$ is in any case close to its equilibrium value,
which does not depend on $g_*^{1/2}$. At temperatures well below the
decoupling temperature, i.e. for $x \gg x_F$, the right--hand side of
the Boltzmann equation (\ref{boltz1}) is suppressed by the explicit
$x^{-2}$ factor. If $\langle \sigma v \rangle \propto 1/x$, as in the
lower frame of fig.~\ref{relicpercentage}, the suppression at $x >
x_F$ is even stronger. Sharp features in $g_*^{1/2}$ therefore give
sharper features, with larger amplitudes, for pure $P-$wave
annihilation than for $S-$wave annihilation.

We noticed earlier that the older treatment of
ref.\cite{Gondolo_Gelmini} overestimates $g_*^{1/2}$ for some range of
temperatures above $T_c$. Eq.(\ref{appsol}) indicates that this should
lead to a smaller predicted relic density, which is confirmed by
fig.~\ref{relicpercentage}. Since $g_*^{1/2}$ is over--estimated for
an extended range of temperatures, the effect on the relic density is
about the same for $S-$ and $P-$wave annihilation, amounting to about
$5\%$ near the peak of the ratio shown in
fig.~\ref{relicpercentage}. The second, much lower peak near $m_\chi
= 1$ TeV is probably due to lack of knowledge of the top mass at the
time when ref.\cite{Gondolo_Gelmini} was written.

The spike in $g_*^{1/2}$ predicted by the {\tt micrOMEGAs} treatment
of the results of ref.\cite{Hindmarsh} gives prominent spikes in the
ratios shown in fig.\ref{relicpercentage}. These spikes are numerical
artefacts that result from the smoothing procedure used in {\tt
  micrOMEGAs}. As argued in ref.\cite{Hindmarsh}, a true $\delta$
function spike in $g_*^{1/2}$ should not affect the numerical solution
of the Boltzmann equation, which necessarily entails some
discretization. The probability that the program then has to evaluate
the right--hand side of the Boltzmann equation at the precise value of
$x$ where the $\delta$ function diverges is zero. We nevertheless show
results including this spike since it results from the ``standard
treatment'' encoded in {\tt micrOMEGAs}. Outside the mass range
affected by this spike, the results of ref.~\cite{Hindmarsh} predict a
slightly too large relic density, consistent with our observation that
it predicts smaller values of $g_*^{1/2}$ and $h$ than our treatment
does. Note that for fixed $g_*^{1/2}$, reducing $h$ will (slightly)
increase $x_F$, leading to an increase of the predicted relic
density. A decrease of $h$ therefore goes into the same direction as a
decrease of $g_*^{1/2}$. However, the fact that the relative
difference between our calculation and the prediction based on
ref.\cite{Hindmarsh} is almost the same in both frames of
fig.~\ref{relicpercentage} at large WIMP masses shows that the main
effect still comes from the change of $g_*^{1/2}$. 

We saw in fig.~\ref{doffig} that the prediction for $g_*^{1/2}$ from
ref.\cite{Laine_Schroeder} lies below our prediction, except for a
very narrow range of temperatures around $T_c$. As a result, for pure
$S-$wave annihilation the prediction for the relic density based on
the treatment of ref.\cite{Laine_Schroeder} lies above our prediction
for all WIMP masses larger than 2 GeV. We argued above that the
relevant range of temperatures is (even) smaller for pure $P-$wave
annihilation. This explains why the blue curve in the lower frame of
fig.~\ref{relicpercentage} goes slightly above 1 for $m_\chi \simeq
25$ GeV. Note also that the predictions using our treatment agrees
with the prediction using ref.\cite{Laine_Schroeder} to better than
$1\%$ for all WIMP masses, except in the range between $3$ and $15$
GeV where the difference reaches $9 \ (12)\%$ for pure $S- \ (P-)$wave
annihilation.

In order to put these results into perspective, it should be noted
that the lattice QCD predictions for the energy and entropy densities
listed in Table 1 of \cite{Bazavov:2014pvz}, on which our treatment is
based, still have significant uncertainties, which decrease from about
$14\%$ at $T = 130$ MeV to about $3 \%$ at $T= 400$ MeV. The
corresponding uncertainty in the relic density is up to $2.5\%$ for
$2$ GeV $\leq m_{\chi}\leq 20$ GeV. Finally, we note that treating the
charm quark as a free particle would increase the predicted relic
density by about $2.2\%$ for $m_\chi \simeq 30$ GeV.

The results shown in fig.~\ref{relicpercentage} are almost independent
of the assumed value of the WIMP annihilation cross section as long as
the relic density comes out at least roughly correctly, although they
evidently are sensitive to the functional dependence of $\langle
\sigma v \rangle$ on the temperature. Eq.(\ref{appsol}) shows that
thermodynamic effects enter primarily through $g_*^{1/2}(x_F)$, and
$x_F$ depends on the annihilation cross section only logarithmically.

On the other hand, the exact value of the annihilation cross section
that reproduces the correct relic density, now (within standard
$\Lambda$CDM cosmology) constrained to be $\Omega_\chi h^2 = 0.1193
\pm 0.0014$ \cite{Planck:2015xua}, does depend on $g_*^{1/2} (T \sim
T_F)$ and, to a lesser extent, on $h(T \sim T_F)$. Precise knowledge
of this cross section is important to constrain the free parameters of
models of thermal WIMPs. Moreover, as we will see in more detail
below, indirect DM searches now begin to probe annihilation cross
sections close to the required value of $\langle \sigma v \rangle$,
{\em if} the latter is (approximately) independent of the temperature.

\begin{figure}[h]   
\centering
\includegraphics[width=13.0cm]{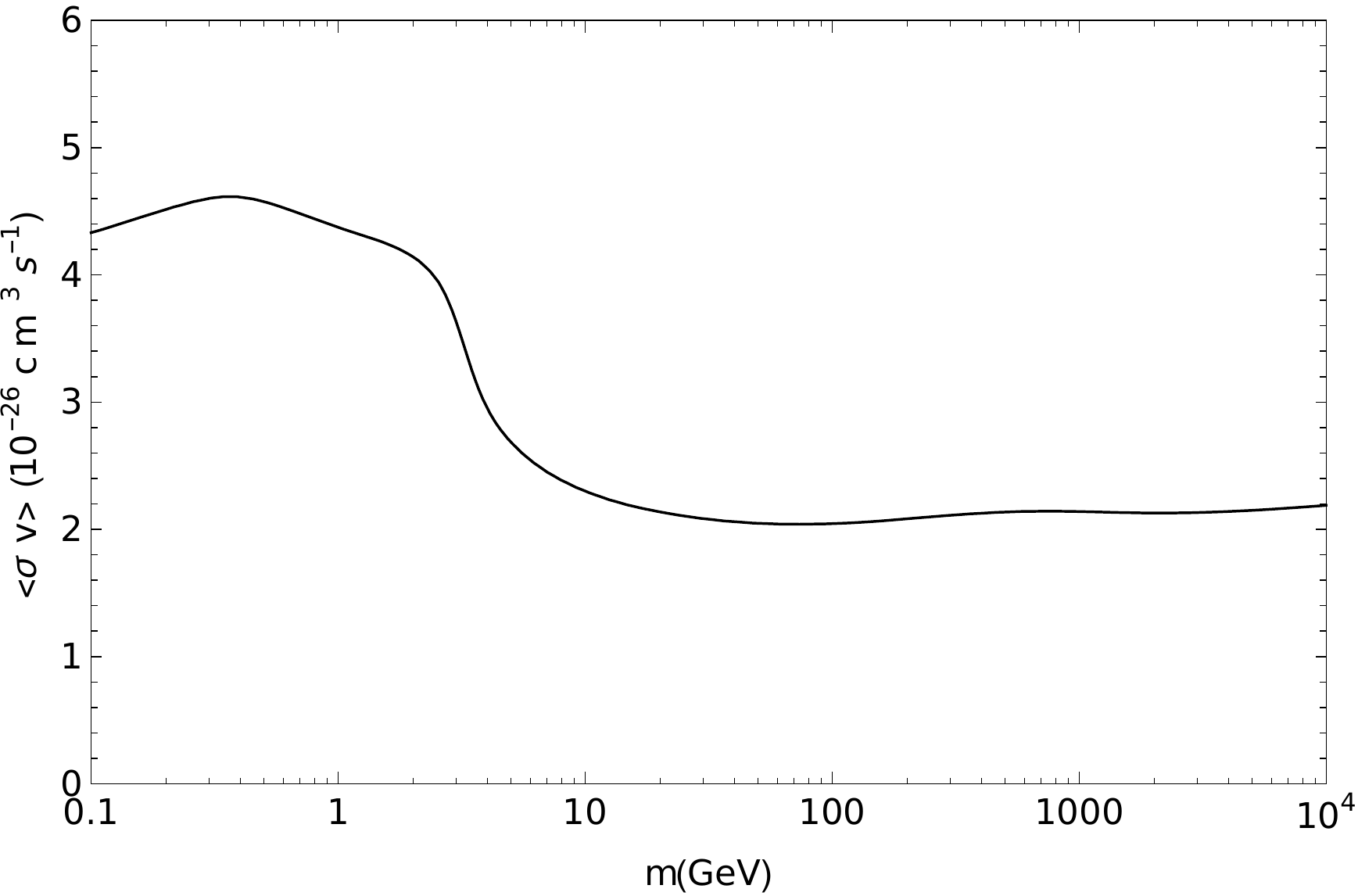}
\caption[] {The value of $\langle \sigma v \rangle$, assumed to be
  completely independent of temperature, required to obtain a thermal
  relic density $\Omega_\chi h^2=0.1193$ within standard cosmology,
  as a function of WIMP mass.}
  \label{newsigmav}
\end{figure}

In fig.~\ref{newsigmav} we show the required value of $\langle \sigma
v\rangle$, assumed to be independent of the temperature, for a Majorana
fermion, obtained from our refined calculation of $g_*^{1/2}$ and
$h$. This updates the results of ref.\cite{Steigman}, which assumed
$\Omega_\chi h^{2}=0.11$ and used \cite{Laine_Schroeder} to compute
$g_*^{1/2}$ and $h$. 

We see that for $10$ TeV $> m_\chi > 10$ GeV the required value of
$\langle \sigma v \rangle$ is in fact closer to $2 \cdot 10^{-26}$
cm$^{3}$s$^{-1}$ than to the ``canonical'', often cited value of $3
\cdot 10^{-26}$ cm$^{3}$s$^{-1}$. The near constancy of the required
value over such a large range of WIMP masses results from an
``accidental'' cancellation of two effects. This can again be
understood from the approximate analytical solution (\ref{appsol}) of
the Boltzmann equation. On the one hand, increasing $m_\chi$ increases
$x_F$, which {\em increases} the relic density. Since $x_F$ depends
only logarithmically on $m_\chi$, the freeze--out temperature $T_F =
m_\chi / x_F$ still increases as $m_\chi$ is increased. As shown in
fig.~\ref{doffig}, this increases $g_*^{1/2}(T_F)$, which in turn {\em
  reduces} the relic density. For $m_\chi > 10$ TeV, all SM particles
are essentially fully relativistic at $T_F$, i.e. $g_*^{1/2}$ becomes
independent of $T$, reaching its asymptotic value of $106.75$. For
these very large WIMP masses the required value of $\langle \sigma v
\rangle$ would thus increase logarithmically with $m_\chi$, in order
to cancel the effect of the increase of $x_F$. However, since by
dimensional analysis and unitarity arguments \cite{uni_lim} $\langle
\sigma v \rangle \ \propto 1/m_\chi^2$, it is very difficult to find
scenarios with sufficiently large WIMP mass for $m_\chi > 10$ TeV.

On the other hand, for WIMP masses below 10 GeV the rapid decrease of
$g_*^{1/2}(T_F)$ with decreasing $T_F$ shown in fig.~\ref{doffig}
requires a rather rapid increase of $\langle \sigma v \rangle$, to a
peak value of about $4.5 \cdot 10^{-26}$ cm$^3$s$^{-1}$. Finally, for
$m_\chi < 0.35$ GeV, $g_*^{1/2}(T_F)$ becomes approximately constant
again, with electrons, positrons, neutrinos, and photons contributing
so that $g_*^{1/2} \simeq 3.29$. Since $x_F$ keeps decreasing with
decreasing $m_\chi$, keeping the relic density constant requires that
$\langle \sigma v \rangle$ also decreases logarithmically with
decreasing WIMP mass for these very light WIMPs.

\section{Experimental constraints on $\langle \sigma v \rangle $}
\label{constraints}

In this section, we will compare experimental constraints from
indirect WIMP searches and from analyses of the cosmic microwave
background (CMB) with our prediction for $\left\langle \sigma v
\right\rangle$ shown in fig.~\ref{newsigmav}. Since the CMB decoupled
much later than WIMPs did, and hence also at a much lower temperature
($\sim 0.3$ eV rather than $\sim m_\chi / 20$), while WIMPs in
galaxies now have an average kinetic energy of $\sim 10^{-6} m_\chi$,
such a comparison is meaningful only if $\langle \sigma v \rangle$ is
largely independent of the temperature. If $\langle \sigma v \rangle
\propto T$, as in pure $P-$wave annihilation, or for even stronger
$T-$dependence, the bounds on $\langle \sigma v \rangle$ from the CMB
and from indirect WIMP searches are still several orders of magnitude
above the value required to obtain the correct relic density.

Currently the strongest and most robust upper bounds on $\langle
\sigma v \rangle$ from indirect WIMP searches come from searches for
hard $\gamma$ rays by the FermiLAT collaboration. Photons travel in
straight lines through our galaxy, whereas charged particles get
deflected by the galactic magnetic field. This not only introduces a
sizable uncertainty, since our knowledge of this magnetic field is far
from perfect; it also isotropizes the arrival directions of the WIMP
annihilation products, making it impossible to focus on regions of
space where the WIMP signal should be particularly strong. Another
advantage of $\gamma$ rays is that they are present in nearly all
possible final states: hard photons can be emitted directly off final
or intermediate state particles, can originate from the decay of
neutral pions and other hadrons that result from the hadronization of
$q \bar q$ final states or the decay of $\tau$ leptons, and can be
produced from energetic electrons or positrons through ``inverse
Compton'' upscattering of ambient photons.

The strongest WIMP signal is expected from near the center of our own
galaxy. Unfortunately this region also hosts several backgrounds, both
in form of point sources and in form of extended emission. It has been
claimed that there is evidence for an additional component in the GeV
$\gamma$ flux from near the galactic center which can be explained
through WIMP annihilation \cite{gc_hooper}, but other interpretations
of this additional component exist \cite{gc_other}. We also note that
the FermiLAT collaboration itself has not published any analysis of
their data on the galactic center.

In this paper we therefore focus on FermiLAT observations of nearby
dwarf galaxies \cite{Fermi-LAT 4,Fermi-LAT 3
  analysis,Ackermann:2013yva, Fermi-LAT new}. In contrast to big
galaxies like our own, the mass density of dwarf galaxies should be
dominated by dark matter even in the central region, yielding a much
better signal--to--background ratio for indirect WIMP signals. No such
signal has been seen. Our analysis is based on the very recent 6--year
``Pass 8'' analysis \cite{Fermi-LAT new}.

\begin{figure}[h]    
\centering
\includegraphics[width=13.0cm]{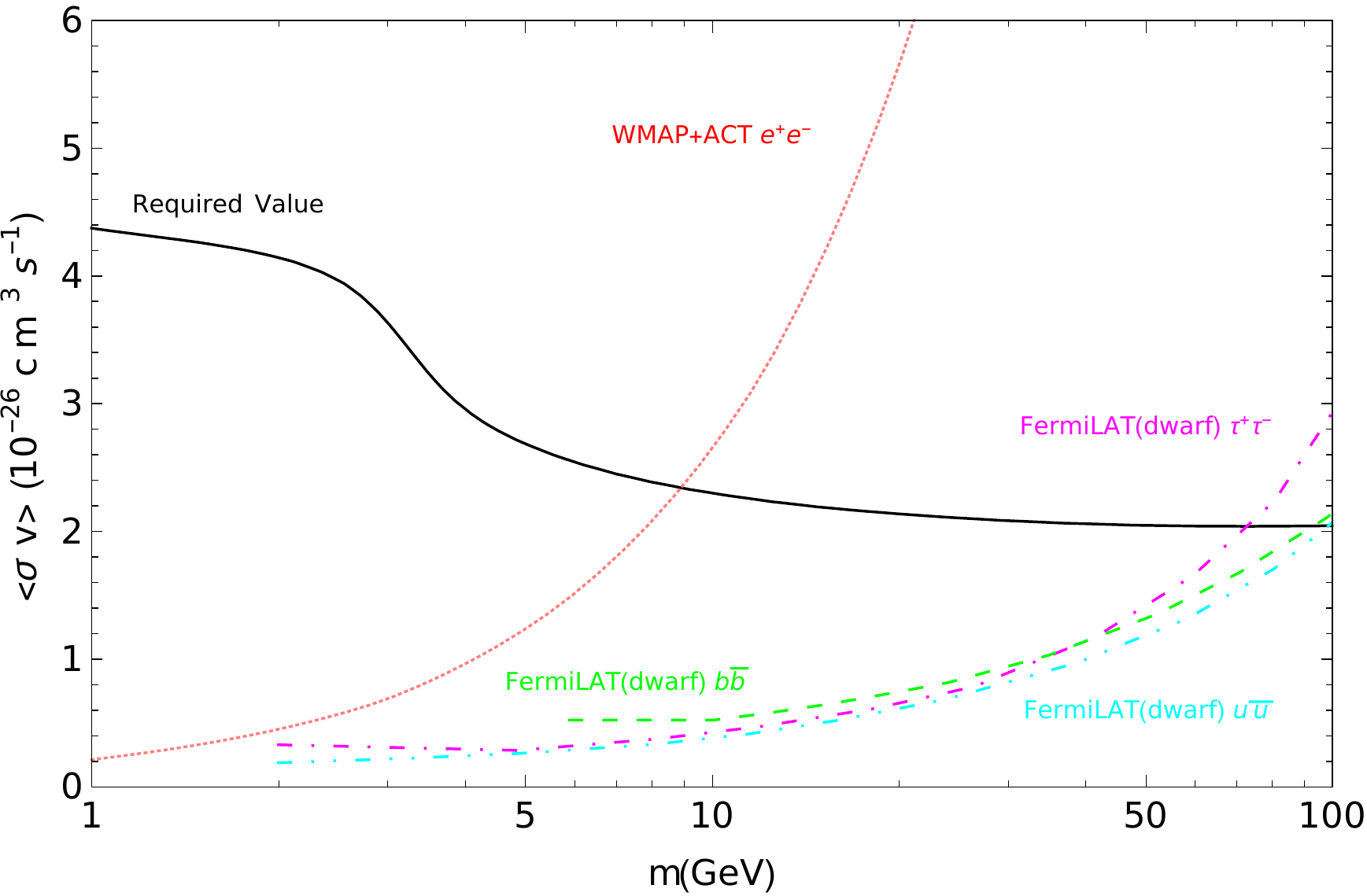}
\caption[] {The result of fig.\,\ref{newsigmav} is compared with
  several observational upper bounds on $\langle \sigma v \rangle$,
  which is assumed to be independent of temperature. The cyan, green
  and magenta curves follow from the FermiLAT upper bound
  \cite{Fermi-LAT new} on the $\gamma$ flux from dwarf galaxies,
  for different dominant WIMP annihilation channel ($u\bar u, \ b \bar
  b$ or $\tau^+\tau^-$), whereas the red curve results from an upper
  bound on spectral distortions of the CMB, assuming WIMP annihilation
  into $e^+e^-$ pairs.}
  \label{newsigmav+experiment}
\end{figure}

The results are shown in fig.~\ref{newsigmav+experiment}. We see that
the upper bound on $\langle \sigma v \rangle$ is strongest if WIMPs
predominantly annihilate into $u \bar u$ final states, but the bound
for WIMP annihilation into $b \bar b$ is only slightly weaker. For the
$\tau^+ \tau^-$ final state the upper bound on the cross section is
similar for WIMP masses below 40 GeV, but is somewhat weaker for
heavier WIMPs; hadronic final states have higher multiplicity, and
hence higher $\gamma$ flux per WIMP annihilation, for larger WIMP
masses, whereas for the $\tau^+\tau^-$ final state the photon
multiplicity is essentially independent of the WIMP mass. These
constraints exclude WIMPs with mass $m_\chi \leq 70$ to $100$ GeV
annihilating into hadrons or $\tau$ leptons with temperature
independent $\langle \sigma v \rangle$.

As mentioned above, WIMP annihilation into $e^+e^-$ can produce hard
photons through upscattering ambient, e.g. visible, photons. The
corresponding upper bound on $\langle \sigma v \rangle$ (not shown) is
worse than that for WIMP annihilation into $\tau^+ \tau^-$ by a factor
of about two to three \cite{Fermi-LAT new}, excluding WIMPs with mass
$m_\chi \leq 15$ GeV annihilating into $e^+e^-$ pairs for the value of
$\langle \sigma v \rangle$ shown in fig.~\ref{newsigmav}.

WIMP annihilation can also affect the CMB. The strongest limits on
$\langle \sigma v \rangle$ originate from the fact that WIMP
annihilation heats up the plasma in the ``recombination'' epoch when
neutral atoms first formed \cite{cmb_old}, thereby delaying the
decoupling of the CMB photons and distorting the pattern of CMB
anisotropies. In fig.~\ref{newsigmav+experiment} we show the bound on
the WIMP annihilation cross section into $e^+e^-$ pairs that results
from an analysis \cite{WMAP+ACT 3} of data from the WMAP and ACT
collaborations. It excludes a thermal WIMP with $m_\chi < 8$
GeV. 

PLANCK data will lead to considerable stronger constraints
\cite{MSS,Planck:2015xua}. Unfortunately these papers only cite upper
bounds on the product of the WIMP annihilation cross section and an
efficiency factor $f_{\rm eff}$ with which the energy of the WIMP
annihilation products is absorbed in the thermal plasma. Using results
from ref.\cite{SPP}, we estimate that the latest PLANCK data exclude
WIMPs with $m_\chi \lsim 40$ GeV annihilating into $e^+e^-$ pairs with
temperature independent cross section; see also the recent analysis
\cite{Steig_new}. 

Since the efficiency factor should be similar for the other final
states considered in fig.~\ref{newsigmav+experiment}, the CMB
constraint should also be similar for all channels. The current CMB
constraint is thus weaker than the bounds derived from the most recent
FermiLAT data if WIMPs mostly annihilate into $q \bar q$ or
$\tau^+\tau^-$ final states, but is stronger for WIMPs annihilating
predominantly into $e^+e^-$. However, one should keep in mind that the
CMB constraint is less direct. It is conceivable that additional
non--standard ingredients to the CMB fit -- e.g., the presence of
sterile neutrinos, a significant running of the spectral index of
inflation, and/or a large contribution from tensor modes -- can
(partly) compensate the distortions caused by early WIMP annihilation,
thereby weakening the constraint on $\langle \sigma v \rangle$. On the
other hand, the constraint derived from the observation of dwarf
galaxies depends on the assumed dark matter distribution
\cite{Fermi-LAT 3 analysis}. In any case, it is encouraging that
recent astrophysical and cosmological observations begin to probe
relatively light thermal WIMPs with temperature independent
annihilation cross section.

\section{Summary and Conclusions}
\label{conclusion}

Using recent lattice QCD results with dynamical quarks for the
equation of state, we have computed the energy and entropy
densities of the SM with emphasis on temperatures around the
deconfinement transition at $T_c=154$ MeV. These results are described
by the functions $g(T)$ and $h(T)$. Of particular relevance for the
calulcation of the relic density of thermal WIMPs is the quantity
$g_*^{1/2}$ defined in eq.(\ref{gstar}), which also depends on the
derivative of $h$ with respect to the temperature. Our results for
these functions can readily be embedded in the public codes computing
the relic density \cite{Gondolo:2004sc,Belanger:2013oya,Arbey:2011zz}.

Our predictions for the WIMP relic density differ from earlier
treatments that relied on phenomenological models or pure glue lattice
QCD calculations. These differences are most pronounced for WIMP
masses between 3 and 15 GeV; they can reach about $9\%$ for a constant
$\langle \sigma v \rangle$, and up to $12\%$ if $\langle \sigma v
\rangle \propto T$ as in pure $P-$wave annihilation away from poles and
thresholds. These differences are partly due to a spurious divergence
in $h(T)$ leading to a divergence in $g_*^{1/2}$, which results in a
sharp spike in the standard implementation. This illustrates the
importance of ensuring that $h(T)$ is continuous everywhere, not just
during the deconfinement transition but also during electroweak
symmetry breaking at $T \simeq 100$ GeV, which we essentially ignored
in our treatment. 

It should be noted that the uncertainties in the recent lattice
calculation we used still translates into an uncertainty of the
predicted relic density of up to $2.5\%$, considerably larger than the
observational uncertainty (at least in the framework of the standard
$\Lambda$CDM cosmology). On the other hand, for $m_\chi > 20$ GeV our
result for the relic density agrees to better than $1\%$ with the best
previous calculation \cite{Laine_Schroeder}. 

We used our improved treatment of the thermodynamics of the very early
universe to update the calculation of the WIMP annihilation cross
section required to reproduce the observed dark matter relic density,
assuming the thermal average $\langle \sigma v \rangle$ to be
independent of temperature. We found that this cross section is indeed
nearly constant for $10 \ {\rm GeV} < m_\chi < 10 \ {\rm TeV}$, thanks
to a fortuitous cancellation between two competing effects; however,
as also pointed out in ref.\cite{Steigman}, the required value is
closer to $2 \cdot 10^{-26}$ cm$^3$s$^{-1}$ than to the often--cited
``canonical'' value of $3 \cdot 10^{-26}$ cm$^3$s$^{-1}$. On the other
hand, for $m_\chi \lsim 3$ GeV the required value of $\langle \sigma v
\rangle$ exceeds $4 \cdot 10^{-26}$ cm$^3$s$^{-1}$. Producing a
sufficiently large annihilation cross section for such light WIMPs
typically requires the introduction of new light ``mediators'' between
the WIMPs and SM particles. If the mediator mass is $\lsim m_\chi/10$,
the contribution of the mediator particles to the entropy and energy
densities should be included, slightly modifying our results for the
required annihilation cross section shown in fig.~\ref{newsigmav}.

We also compared the required value of $\langle \sigma v \rangle$ with
upper bounds on this quantity that come from searches for energetic
$\gamma$ rays produced in WIMP annihilation as well as from CMB
constraints. The most recent PLANCK data most likely give the
strongest upper bounds, although a dedicated analysis for specific
final states has not yet been performed. Moreover, the CMB bound
assumes standard $\Lambda$CDM cosmology, and can thus more easily be
evaded than the bounds from $\gamma$ ray searches. All of these bounds
can be evaded if the WIMPs predominantly annihilate into neutrinos
\cite{evasion}. 

It should be noted that both our calculation of the required WIMP
annihilation cross section and our analysis of observational upper
bounds on this quantity assumed that the thermal average $\langle
\sigma v \rangle$ is independent of temperature, or, equivalently,
that the annihilation cross section is independent of the invariant
center--of--mass energy. Theoretically this assumption is not well
motivated. For non--relativistic WIMPs the annihilation cross section
can usually (but not always \cite{gs}) be expanded in terms of the
relative velocity $v$, $\sigma v = a + b v^2 + \cdots$. Even if the
constant term $a$ is not suppressed, i.e. if annihilation from an
$S-$wave is allowed, one would generically expect $b$ to be of the
same order as $a$. This would reduce the required value of $\langle
\sigma v \rangle$ in today's universe by about $10\%$. If the constant
term in $\sigma v$ is suppressed, the upper bounds on $\langle \sigma
v \rangle$ that follow from observations in today's universe do not
constrain thermal WIMP models significantly.

However, even in this case it is important to calculate the relic
density as accurately as possible. This leads to a constraint on the
free parameters of the underlying WIMP model, which can hopefully one
day be compared to direct measurements at colliders. Only then will we
be able to say with some confidence that this WIMP model is indeed
correct. This paper makes a small contribution to this ambitious
long--term program.


\section{Acknowledgements}
The authors would like to thank Raghuveer Garani and Masaki Asano for
useful discussions, and Basudeb Dasgupta for clarification of his
work. This work was partially supported by the Deutsche
Forschungsgemeinschaft (DFG) via the Collaborative Research Center
TR--33 ``The Dark Universe'', and partly by the Helmholtz Alliance
Astroparticle Physics.  FH is supported by the Deutsche Akademische
Austauschdienst (DAAD). ERS is supported by Conselho Nacional de
Desenvolvimento Cient\'ifico e Tecnol\'ogico (CNPq).

\end{document}